\documentstyle[12pt,epsf,float,cite]{article}

\newcommand{\BEQ}{\begin{equation}}     
\newcommand{\BEA}{\begin{eqnarray}}
\newcommand{\EEQ}{\end{equation}}       
\newcommand{\EEA}{\end{eqnarray}}
\newcommand{\D}{{\rm d}}                
\newcommand{\II}{{\rm i}}               

\catcode`\@=11
\def\numberbysection{\@addtoreset{equation}{section}
        \def\theequation{\thesection.\arabic{equation}}}

\begin{document}

\begin{titlepage}

\vskip 1.5 cm
\begin{center}
{\Large \bf Exact solution of a reaction-diffusion process with three-site
interactions}
\end{center}

\vskip 2.0 cm
\centerline{  {\bf Malte Henkel$^{a,b}$ and Haye Hinrichsen$^{c}$}}
\vskip 0.5 cm
\centerline {$^a$Laboratoire de Physique des 
Mat\'eriaux,\footnote{Laboratoire associ\'e au CNRS UMR 7556} 
Universit\'e Henri Poincar\'e Nancy I,} 
\centerline{ B.P. 239, 
F -- 54506 Vand{\oe}uvre-l\`es-Nancy Cedex, France\footnote{permanent address}}
\centerline{$^{b}$Complexo Interdisciplinar, Faculdade de Ci\^encias, 
Universidade de Lisboa, }
\centerline{Av. Prof. Gama Pinto 2, P -- 1649-003 Lisboa, Portugal}
\centerline{$^c$Theoretische Physik, Fachbereich 10, 
Gerhard-Mercator Universit\"at Duisburg,}
\centerline{ D -- 47048 Duisburg, Germany}
\begin{abstract}
\noindent 
The one-dimensional
reaction diffusion process $AA\to A$ and $A\emptyset A\to AAA$ is
exactly solvable through the empty interval method if the diffusion rate equals
the coagulation rate. Independently of the particle production rate, the
model is always in the universality class of diffusion-annihilation. This
allows us to check analytically the universality of finite-size scaling in
a non-equilibrium critical point.  
\end{abstract}
\end{titlepage}

\section{Introduction}

The physics of non-equilibrium phase transitions in low dimensions is
characterized by the presence of strong fluctuation effects which modify the
properties of the steady state and/or the long-time behaviour considerably
with respect to simple kinetic equations, 
see \cite{Priv96,Chop98,Marr99,Schu00,Hinr00} for recent reviews. However,
and in contrast to equilibrium phase transition, most of the current 
understanding of non-equilibrium phase transitions comes from numerical
studies of certain simple models. Integrable non-equilibrium systems are still
very rare. 

A model which has met recently with a lot of interest 
is the one-dimensional pair contact
process \cite{Jens93} with single particle diffusion (PCPD). This model 
describes the interactions of a single species $A$ of particles which can
undergo the reactions $2A\to\emptyset$ and $AA\emptyset\to AAA$. If the
diffusion constant $d=0$, there is a huge number of steady states 
(for $L$ sites of the order $\sim\phi^L$, where $\phi=(1+\sqrt{5})/2$ is the 
golden mean \cite{Carl00}) and the steady state phase transition between the 
active and the inactive phases is in the directed percolation universality 
class \cite{Jens93} (for $d=0$, the mean pair density serves as the order
parameter which is positive in the active phase and vanishes at the transition
towards to inactive phase). On the other hand, 
for $d\ne 0$, there remain just two 
steady states. There is a general agreement on the location of the transition
line between the active and the inactive phases (current studies apply either
the density matrix renormalisation group \cite{Carl00} or different Monte
Carlo schemes \cite{Hinr00a,Odor00}) and that the entire inactive phase should
be critical and in the diffusion-annihilation universality class \cite{Howa97}.
However, so far there is no consensus 
on the exponents {\em at\/} the active-inactive transition. 

In order to get a fresh view on this problem, we try to find a `nearby' model
where some analytical information might be available. In this paper, we shall
study the following model: particles of a single species move along a 
one-dimensional lattice. Each site can be either empty ($\emptyset$) or else
be occupied by a single particle ($A$). Between nearest neighbour or 
next-nearest neighbour sites the following reactions are allowed
\BEQ \label{gl:Raten}
\begin{array}{lll}
\mbox{diffusion:} &
A \emptyset \leftrightarrow \emptyset A  & \mbox{\rm ~; rate $d$} \\
\mbox{coagulation to the left:} \ \ \ \ \ &
A A \to A \emptyset                      & \mbox{\rm ~; rate $d$} \\
\mbox{coagulation to the right:}  &
A A \to \emptyset A                      & \mbox{\rm ~; rate $d$} \\
\mbox{production:} &
A \emptyset A \to A A A                  & \mbox{\rm ~; rate $2d\lambda$}
\end{array}
\EEQ
The equality of the diffusion rate and the
coagulation rates guarantees the integrability of the
model for $\lambda=0$ through the empty interval method~\cite{Avra90}.

As will be explained in the following section,
the choice of this model is motivated by the 
equivalence of simple coagulation and annihilation. 
Considering the above model on a finite lattice with $L$ sites and periodic
boundary conditions, we shall show in section 3 that it is integrable through 
the finite-size empty interval method \cite{Kreb95}. We obtain explicitly the
long-time behaviour of the particle density and study the finite-size scaling
of the leading relaxation time $\tau$. We find that for all values of $\lambda$,
the model remains in the universality class of pair annihilation and we 
discuss the physical reasons for this. While that
does not provide any insight into the phase transition of the
pair contact process, the fact that $\lambda$ couples to an irrelevant 
operator allows us to test analytically the generalisation \cite{Henk00} 
of Privman-Fisher universality of finite-size scaling 
amplitudes \cite{Priv84} in an exactly solvable non-equilibrium model.
Section 4 summarizes our conclusions.

\section{Motivation of the model}

Although the present understanding of the phase transition in the PCPD
is far from being complete, it is near at hand to conjecture that the 
same type of transition occurs in a large variety of other models.
More precisely, we expect such transitions to occur
in models without parity
conservation where (a) solitary particles diffuse,
(b) particle creation requires two particles and (c) particle
removal requires at least two particles to meet
at neighboring sites.
Examples of such reaction-diffusion processes include
\begin{eqnarray}
2A \rightarrow 3A, & 2A \rightarrow A \label{CoagFiss} \\
2A \rightarrow 4A, & 2A \rightarrow A \\
2A \rightarrow 3A, & 3A \rightarrow \emptyset \\
2A \rightarrow 3A, & 3A \rightarrow A 
\end{eqnarray}
%
%
\begin{figure}
\epsfxsize=120mm
\centerline{\epsffile{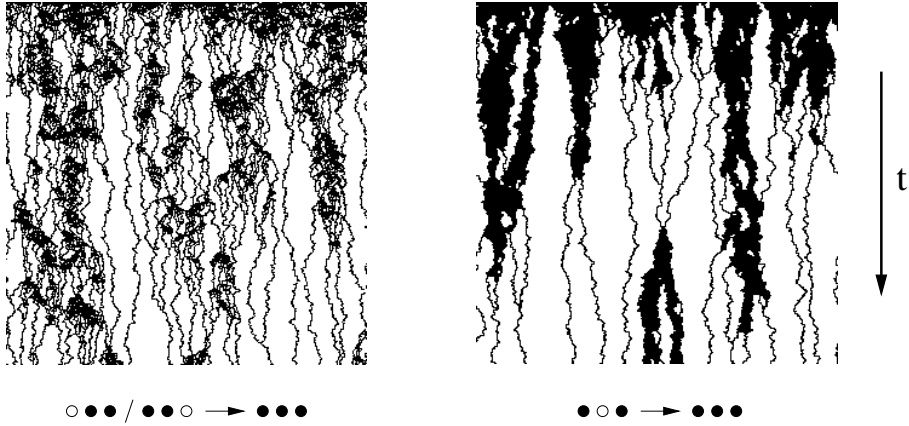}}
\caption{\footnotesize
Typical temporal evolution of the coagulation-production process.
Left: If particles are created to the left and to the right of a pair
the system displays a phase transition similar to the one observed
in the PCPD. Right: If particles are created {\it between} two
particles the model is always in the inactive phase. The figure
shows a typical run for $\lambda=100$.}
\end{figure}
%
%
A particularly interesting candidate is the coagulation/production process
(\ref{CoagFiss}). It is well known that coagulation
$2A\to A$ and pair annihilation $2A\to\emptyset$ are equivalent
and can be related by an exact similarity transformation
(for reviews see~\cite{Schu00,Hinr00}). Assuming that this transformation
does not entirely destroy the production process $2A \to 3A$
in the renormalization group sense, it is therefore natural 
to expect that the coagulation/production
model exhibits the same  type of phase transition as the PCPD.

It is important to note that the production process $2A \to 3A$ 
in one dimension can
be implemented in two different ways. In the standard implementation
particles are created to the left and to the right of a pair 
of particles:
\begin{equation}
\label{OrdinaryFission}
\bullet\bullet\circ/\circ\bullet \ \bullet\to\bullet\bullet\bullet\,.
\end{equation}
Together with the coagulation process this implementation
of the model displays a non-equilibrium phase transition 
similar to the transition in the PCPD (see left panel of Fig.~1).
In the other implementation of the production process,
on which we will focus in the present work,
a particle is created {\it between} two other particles:
\begin{equation}
\label{MiddleFission}
\bullet\circ\bullet\to\bullet\bullet\bullet\,.
\end{equation}
In this case the model displays a completely different behaviour.
In particular, there is no phase transition. Rather the model
is always in the inactive phase where the asymptotic behaviour
is governed by the coagulation process. 
Even the visual appearance of clusters is clearly 
different in both cases, as demonstrated in Fig.~1.
Obviously, the spatial arrangement of the production process
is crucial on a one-dimensional chain. 
Similar hard-core effects can be observed in other 
one-dimensional models with three-site 
interactions~\cite{Meno97}.

\begin{figure}
\epsfxsize=80mm
\centerline{\epsffile{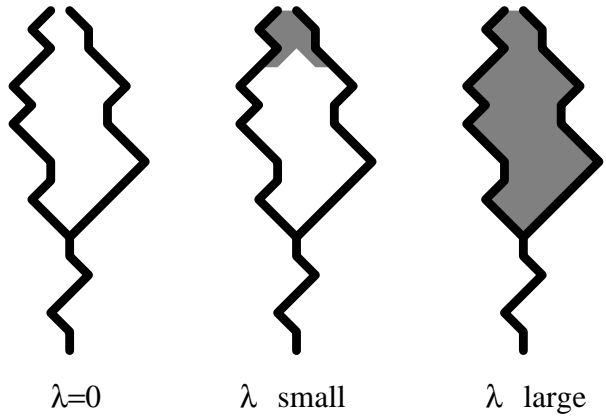}}
\caption{\footnotesize
Schematic drawing of the temporal evolution 
of a pair of particles. For $\lambda=0$ the
two particles diffuse until they coagulate, forming a skeleton
(bold lines). For $\lambda>0$ particles are created between pairs
of particles, forming a high-density region between the two branches.
The skeleton itself remains unchanged.}
\end{figure}

Fig.~2 illustrates why particle production between particles
differs significantly from ordinary particle creation to the
left and to the right. The figure sketches the temporal evolution
of a pair of particles for a given realization of randomness.
Without offspring production, the two particles diffuse until
they meet and coagulate to a single particle. Adding the process
(\ref{MiddleFission}), particles can only be created {\it between}
the space-time trajectories of these two particles. In other
words, the pure coagulation process provides a `skeleton'. The
process (\ref{MiddleFission}) generates additional patches
of high activity between two branches
while the skeleton itself remains unchanged.
Moreover, there is no way for the particles to `cross' the 
branches of the skeleton.
Therefore, the asymptotic behaviour for $t \to \infty$ is governed
by the pure coagulation process, i.e., we expect the density
of particles to decay as $t^{-1/2}$. 

As can be seen in the right panel of Fig.~1, even for high values
of $\lambda$ most patches with high activity die out after short time.
New patches with finite life time can only be generated if two
diffusing particles meet. As these events become more and more 
rare as time proceeds,
it seems to be plausible (and will be proven below) that the model 
does not exhibit a phase transition.  

Let us point out that the PCPD with production in the middle 
$A \emptyset A \to AAA$ and pair annihilation 
$AA \rightarrow \emptyset \emptyset$ 
does not have these special properties. Moreover, it is
not exactly solvable and exhibits a phase transition similar
to the standard PCPD.

\section{Exact solution}

We consider the model (\ref{gl:Raten}) on a chain of $L$ sites with periodic
boundary conditions. The exact solution can be obtained through the 
generalisation of the empty-interval method \cite{Avra90,Pesc94} to finite 
lattices \cite{Kreb95}. Since we are interested in working out spatially 
averaged values of the observables, 
we can assume translation invariance from the
outset. Let $I_n(t)$ be the probability that $n$ consecutive sites are empty
at time $t$. Then the mean particle density is given by
\BEQ \label{IDichte}
\rho(t) = \left( 1 - I_1(t) \right) a^{-1} \,,
\EEQ
where $a$ is the lattice constant. The equations of motion for the $I_n(t)$ are
\BEA
I_0(t)       &=& 1 \nonumber \\[2mm]
\frac{\D I_1}{\D t} (t) 
             &=& 2d \left[ I_0(t) -2I_1(t) + I_2(t) \right] 
                -2d\lambda \left[ I_1(t) -2 I_2(t) + I_3(t) \right]
             \nonumber \\[2mm]
\frac{\D I_n}{\D t}(t) 
             &=& 2d \left[ I_{n-1}(t) -2I_n(t) + I_{n+1}(t) \right] 
                 \;\; ; \;\; ~~~~~~~~~2\leq n\leq L-1 \nonumber \\[2mm]
I_L(t)       &=& 0  \label{gl:gl}
\EEA
Here the boundary condition $I_0(t)=1$ allows one to take care
of the coagulation process in the usual way, 
provided that the rates for coagulation and
diffusion coincide \cite{Avra90,Kreb95}. Obviously, the empty lattice
\BEQ
I_n(t)=1
\EEQ
is a trivial stationary state which decouples from all other solutions.
Therefore, we restrict our analysis to solutions with at least one particle.
Since the last particle can never disappear, 
it is impossible to have $L$ consecutive empty sites for
a chain with $L$ sites, leading to the other boundary condition $I_L(t)=0$.
For $\lambda=0$, eqs. (\ref{gl:gl}) reduce to the known ones for simple
coagulation with periodic boundary conditions \cite{Kreb95}. 

In order to understand the $\lambda$-dependent term,
consider the probability $P(n_1 n_2 n_3)$ of realising the configuration
$n_1 n_2 n_3$ at three neighbouring sites, where $n=\bullet$ indicates an
occupied and $n=\circ$ an empty site. In particular, we have
$P(\circ\circ\circ)= I_3$ and 
$P(\bullet\circ\circ) = P(\circ\circ\bullet) = I_2 - I_3$. In addition,
summing over the states of the third site
\BEQ
P(\bullet\circ\bullet) + P(\bullet\circ\circ) = P(\bullet\circ) = I_1 - I_2
\EEQ
which yields $P(\bullet\circ\bullet)=I_1 -2I_2 + I_3$. The production of a
particle between two others via the process
$\bullet\circ\bullet\to\bullet\bullet\bullet$ only affects $I_1(t)$.

The equations of motion are solved through the ansatz
\BEQ
I_n(t) = \sum_{\omega} a_n(\omega) e^{-2d\omega t}
\EEQ 
which leads to the eigenvalue problem
\begin{displaymath}
\left[ 
\begin{array}{cccccccc}
0      & 0          & 0          & \cdots   &   &        &    & 0 \\
1      & -2-\lambda & 1+2\lambda & -\lambda & 0 & \cdots &    & 0 \\
0      & 1          & -2         & 1        & 0 & \cdots &    & 0 \\
       & 0          & 1          & -2       & 1 & 0      & \cdots & 0 \\
\vdots &            & \ddots     & \ddots   & \ddots & \ddots &\ddots & \vdots\\
       &            &            & 0        & 1 & -2     & 1  & 0 \\
       &            &            &          & 0 & 1      & -2 & 1 \\ 
0      & \cdots     &            &          &   & 0      & 0  & 0  
\end{array} \right]
\left[ \begin{array}{c}
a_0 \\ a_1 \\ a_2 \\ ~ \\ \vdots \\ ~ \\ a_{L-1} \\ a_L \end{array} \right]
= - \omega 
\left[ \begin{array}{c}
a_0 \\ a_1 \\ a_2 \\ ~ \\ \vdots \\ ~ \\ a_{L-1} \\ a_L \end{array} \right]
\end{displaymath} 
involving an $(L+1)\times(L+1)$ matrix $\hat{\Omega}$. 
The irreversible character of the stochastic process reflects itself in the 
fact that the matrix $\hat{\Omega}$ is not symmetric, while probability
conservation implies that the sum of the elements in a
row of $\hat{\Omega}$ vanishes. Because of these properties, 
the real part of the eigenvalues $\omega$ is non-negative. 

It is easy to see that the solution
\BEQ
\label{StationaryState}
a_n(0) = 1 - n/L \;\; ; \;\; \omega = 0
\EEQ
describes the steady state with a single diffusing particle. 
Therefore, the model has {\em two} steady states,
one corresponding to the empty lattice and the other one being the 
translation-invariant superposition of all single-particle states with an
average density $\rho_{\rm av}=1/L$. 
For the relaxational modes with $\omega > 0$
the boundary condition $I_0(t)=1$ implies that $a_0(\omega)=0$.
Similarly, the other boundary condition
$I_L(t)=0$ implies $a_L(\omega)=0$. Therefore, we go over to an
eigenvalue problem involving an $(L-1)\times(L-1)$ matrix if $\omega\ne 0$
\BEQ \label{gl:Eigen}
\left[ 
\begin{array}{ccccccc}
 -2-\lambda & 1+2\lambda & -\lambda & 0          & \cdots &        & 0     \\
 1          & -2         & 1        & 0          & \cdots &        & 0     \\
 0          & 1          & -2       & 1          & 0      &        &       \\
 \vdots     & \ddots     & \ddots   & \ddots     & \ddots &\ddots  & \vdots\\
            &            & 0        & 1          & -2     & 1      & 0     \\
            &            &          & 0          & 1      & -2     & 1     \\
 0          & \cdots     &          &            & 0      & 1      & -2  
\end{array} \right]
\left[ \begin{array}{c}
a_1 \\ a_2 \\ ~ \\ \vdots \\ ~ \\ a_{L-2} \\ a_{L-1} \end{array} \right]
= - \omega 
\left[ \begin{array}{c}
a_1 \\ a_2 \\ ~ \\ \vdots \\ ~ \\ a_{L-2} \\ a_{L-1} \end{array} \right]\,.
\EEQ
Eq. (\ref{gl:Eigen}) is solved through the ansatz
\BEQ \label{gl:Ansatz}
a_n(\omega) = A e^{\II k n} + B e^{-\II k n} \;\; ; \;\; \omega\ne 0
\EEQ
which leads to the dispersion relation 
\BEQ
\omega=\omega(k) = 2(1-\cos k)
\EEQ 
and the allowed values $k$ are obtained by inserting the ansatz 
(\ref{gl:Ansatz}) into the first line of eq.~(\ref{gl:Eigen}) and taking
the boundary condition $a_L(\omega)=0$ into account. This leads to a
system of two equations
\BEA
A \left( \lambda \left( e^{\II k} -2e^{2\II k}+e^{3\II k}\right) +1 \right)
+ B \left( \lambda \left( e^{-\II k} -2e^{-2\II k}+e^{-3\II k}\right) +1 \right)
&=& 0 \nonumber \\
A e^{\II k L} + B e^{-\II k L} &=& 0 \nonumber 
\EEA
which has a non-trivial solution if $k$ is a solution of 
\BEQ \label{gl:Disk}
\tan kL = 
\frac{4\lambda\sin(2k) \sin^2(k/2)}{4\lambda\cos(2k) \sin^2(k/2)-1} \,.
\EEQ
We call the solutions of (\ref{gl:Disk}) $k_m$, where $m=1,\ldots,L-1$. 
In addition, we can include the stationary solution by letting $k_0=0$. 
Having found these, we can write the final result for the empty interval 
probabilities $I_n(t)$ in the form
\BEQ \label{gl:Ende}
I_n(t) = \left( 1 - \frac{n}{L}\right) 
+ \sum_{m=0}^{L-1} C_m \sin\left( k_m (n-L)\right) e^{-2d\,\omega_m\, t}\,,
\EEQ
where $\omega_m = \omega(k_m) = 2(1-\cos k_m)$ and the $C_m$ are real
constants which must be determined from the initial conditions.  
For example, for an initially fully occupied lattice, one has
$I_n(0)=\delta_{n,0}$. If we insert the values of $k_m$ into (\ref{gl:Ende})
and use (\ref{IDichte}), we obtain the average particle density as a function
of time. 

Closed-form solutions of (\ref{gl:Disk}) exist for $\lambda=0$ and 
$\lambda\to\infty$. We find $k_m= m\pi/L$ for $\lambda=0$ and
$k_m= m\pi/(L-2)$ for $\lambda\to\infty$, respectively. 
For general values of $\lambda$, we have
\BEQ \label{kWerte}
k_m = \frac{m\pi}{L} - \frac{2\pi^3 \lambda m^3}{L^4} + \ldots
\;\; ; \;\; m =0,1,\ldots,L-1
\EEQ
Since the asymptotic scaling of the $k_{m}\sim L^{-1}$ is the same for
$\lambda$ finite and for $\lambda=\infty$, even the point $\lambda=\infty$
cannot be interpreted as being a transition point towards a different 
phase. 

{}From eqs.~(\ref{gl:Ende},\ref{kWerte})
we see that the {\em exact\/} inverse leading relaxation time $\tau$ is given by
\BEQ \label{gl:tauUniv}
\tau^{-1} = 2 d \omega(k_1) \simeq 2 d \pi^2 L^{-2} 
\left( 1 + O\left(L^{-2}\right)\right)\,.
\EEQ
In other words, the finite-size scaling amplitude 
\BEQ
A := \lim_{L\to\infty} L^2 \tau_L^{-1} = 2 d \pi^2
\EEQ
is independent of the 
particle production rate $\lambda$, confirming the 
hand-waving arguments presented in Sect. 2. 
We point out that the value of $A$ is
equal to the value of the finite-size scaling amplitude of the
leading relaxation time in the entire inactive
phase in the pair contact process \cite{Henk00}. 
An analogous universality holds for the entire spectrum of relaxation
times $\tau_{m}^{-1} = 2d \omega(k_m)$. 

%
%
\begin{figure}
\epsfxsize=80mm
\centerline{\epsffile{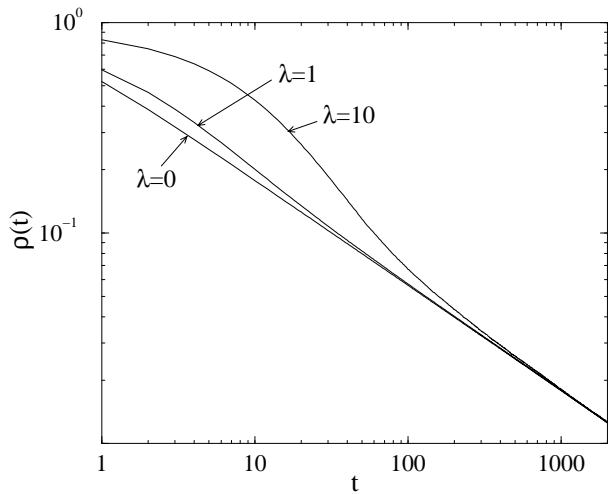}}
\caption{
\footnotesize
\label{FigDecay}
Particle density as a function of time for
various values of $\lambda$. All curves converge
to a single one, demonstrating the irrelevance
of the parameter $\lambda$.}
\end{figure}
%
%

Starting with a fully occupied lattice, the leading relaxation time 
is proportional to the time needed to reach the steady state
(\ref{StationaryState}) with density $\rho(\infty)=1/L$, 
see eq.~(\ref{gl:Ende}). 
As an immediate consequence (see section 4), 
the asymptotic decay of the particle density
has to be independent of $\lambda$ as well. To verify this prediction,
we performed Monte Carlo simulations. As shown in Fig.~\ref{FigDecay},
the production process affects the curves only on a limited time window.
Eventually all curves converge, demonstrating the universality of the
long-time behaviour with respect to $\lambda$.

\section{Conclusions}

We have seen that in our integrable 
coagulation-production model (\ref{CoagFiss}),
the finite-size scaling of the relaxations times $\tau_m$ and of the
steady-state particle density~$\rho$ is independent of $\lambda$. We have
also shown how to understand this in a physical way. In addition, this
result can also be understood in the context of a recent extension \cite{Henk00}
of the Privman-Fisher scaling forms \cite{Priv84} (see \cite{Priv91} for a 
review) to the steady states
of non-equilibrium phase transitions below their upper critical dimension. 
In particular, for a $1D$ reaction-diffusion system of finite length $L$, 
the relaxation times $\tau_m$ should scale as (using the same notation as in
\cite{Henk00})
\BEQ
\tau_{m}^{-1} = C_0 L^{-z} 
\left. R_m\left( 0, C_2 h L^{1+z-\beta/\nu_{\perp}}\right)\right|_{h=0}
\EEQ
and the steady-state particle density as
\BEQ
\rho = C_2 L^{-\beta/\nu_{\perp}} 
\left. Y'\left( 0, C_2 h L^{1+z-\beta/\nu_{\perp}}\right)\right|_{h=0}
\EEQ
where $\beta,\nu_{\perp},z$ are the order parameter, correlation length and
dynamical exponents, $R_m$ and $Y'$ are universal scaling functions, $h$
parametrizes an external source of particles, and $C_0, C_2$ are non-universal
constants \cite{Henk00}. For the diffusion-coagulation model at hand 
(when $\lambda=0$), it is known that $z=2$ and that the time scale can be
fixed by choosing $C_0=d$ \cite{Droz93}. The $\lambda$-independence of all
the $\tau_m$ is therefore consistent with the expected universality of the
$R_m$. Furthermore, the steady-state density is given by
$\rho(\infty)=1/L$, which implies that even the generically non-universal
constant $C_2$ does not depend on $\lambda$ and that $\beta/\nu_{\perp}=1$. 
A simple scaling argument then shows that, in the limit $L\to\infty$, also
the time-dependent density
\BEQ
\rho(t) = \sum_m \rho^{(m)} e^{-t/\tau_m} \mathop{\simeq}_{t\to\infty}
\rho_0 (d t)^{-\delta}
\EEQ
where $\delta=\beta/\nu_{\|}=\beta/(\nu_{\perp}z)=1/2$ and the constant
$\rho_0$ is $\lambda$-independent. That is indeed what we observe from 
figure~\ref{FigDecay}.

Recall that in the pair contact process similar arguments demonstrate
that $C_0$ is not renormalized through the effects of the interactions,
see \cite{Howa97,Henk00} and references therein. However, the
universality of the amplitude $A$ in the pair contact process could only
be established numerically \cite{Henk00}. On the other hand, the universality
of the relaxation times in the annihilation-coagulation model
$2A\to \emptyset, A$ is trivial because of a similarity transformation which
reduces the model to simple diffusion-annihilation. We have therefore obtained 
the first non-trivial analytic confirmation of the universality of the 
finite-size scaling amplitudes of the correlation length.

\newpage
\noindent {\large\bf Acknowledgements}\\

\noindent
We thank J-M. Luck and C. Godr\`eche for organising the stimulating ambiance
of the 5$^{\mbox{\rm\small \`emes}}$ Rencontres C. Itzykson, 
where this work was started. 
MH thanks the Complexo Interdisciplinar of the University of Lisbon for warm
hospitality, where this work was done. 



\end{document}